\documentclass[preprint]{elsarticle}
\usepackage[utf8]{inputenc}
\usepackage{array,amssymb,amsmath}
\usepackage{graphics,graphicx}
\usepackage{natbib}
\usepackage{aas_macros}

\def\int {\emph{INTEGRAL}}

\begin{document}
\begin{frontmatter}

\title{\int\ Results on Gamma-Ray Bursts and Polarization of Hard X-ray Sources}

\author{Diego G\"otz\corref{cor1}}%
\ead{diego.gotz@cea.fr}

\author{Christian Gouiff\`es, J\'er\^ome Rodriguez}

\address{AIM, CEA, CNRS, Universit\'e Paris-Saclay, Universit\'e Paris Diderot, Sorbonne Paris Cit\'e, F-91191 Gif-sur-Yvette, France}

\author{Philippe Laurent}

\address{Laboratoire APC, 10, rue Alice Domon et L\'eonie Duquet, F-75205 Paris Cedex 13, France}

\author{Elisabeth Jourdain, Jean-Pierre Roques}
\address{CNRS; IRAP; 9 Av. colonel Roche, BP 44346, F-31028 Toulouse cedex 4, France\\
 Universit\'e de Toulouse; UPS-OMP; IRAP;  Toulouse, France\\}

\author{Sandro Mereghetti}

\address{INAF - Istituto di Astrofisica Spaziale e Fisica Cosmica Milano, Via E. Bassini 15, 20133, Milano, Italy}

\author{Alexander Lutovinov}

\address{Space Research Institute, Profsoyuznaya str. 84/32, 117997 Moscow, Russia}

\author{Volodymyr Savchenko}

\address{ISDC/University of Geneva, chemin d'\'Ecogia, 16- 1278 Versoix, Switzerland}

\author{Lorraine Hanlon, Antonio Martin-Carrillo}

\address{Space Science Group, School of Physics, University College Dublin, Belfield, Dublin 4, Ireland}

\author{Paul Moran}

\address{Centre for Astronomy, School of Physics, National University of Ireland, H91 TK33 Galway, Ireland}

\begin{abstract}
In this paper we first review the results obtained by the \int\ mission in the domain of Gamma-Ray Bursts (GRBs), thanks to the \int\ Burst Alert System, which is able to deliver near real-time alerts for GRBs detected within the IBIS field of view. More than 120 GRBs have been detected to date and we summarize their properties here. In the second part of this review we  focus on the polarimetric results obtained by IBIS and SPI on GRBs and Galactic compact objects.
\end{abstract}


\end{frontmatter}


\section{Introduction}
Although the \int\ mission was not originally designed as a Gamma-Ray Bursts (GRBs) oriented mission, thanks to the big field of view of its instruments, its excellent sensitivity in the soft $\gamma$-ray domain and its good localization capabilities, it has been used during its entire lifetime as a real-time GRB localization instrument, and an almost-all-sky transient source detector. This allowed \int\ to provide some significant contribution to GRB science, that will be summarized in Section \ref{sec:grbs}.

Similarly to GRB science, hard-X/soft-$\gamma$-ray polarization results were obtained using IBIS \citep{ibis} and SPI \citep{spi}, although both telescopes were not explicitly designed as polarimeters. Polarization results obtained by \int\ during the last years will be presented in Section \ref{sec:grb_pola}, for what concerns GRBs, and in Section \ref{sec:pola} for what concerns other compact sources, such as the Crab and black hole binaries.


\section{Gamma-Ray Bursts}
\label{sec:grbs}

The capabilities of \int\  for GRB studies were recognized early during the development phase of the mission and led to the development of a system for the rapid localization and public distribution of the relevant information:  the \int\  Burst Alert System (IBAS, \cite{ibas03}).   

The IBIS instrument can provide localizations with arcmin accuracy for the bursts revealed in its imaging field of view of $\sim$30x30 deg$^2$, while the large collecting area of the SPI Anti Coincidence Shield allows the measurement  of  light curves of GRBs in almost any position of the sky.   All the IBAS software runs at the \int\  Science Data Center (ISDC), exploiting the continuous telemetry link that allows a very short latence time.   In this respect, \int\  is different from all other $\gamma$-ray satellites used for GRB studies, because its instruments do not have  on-board triggering capabilities and/or specific operating modes for transient events.  

The IBAS GRB localization is based on  imaging data from the IBIS/ISGRI detector. These data  have an average  latency of $\sim$5 s on board the satellite, to which a short delay (typically about 6 s) must be added before they reach the ISDC, where they are immediately extracted from the telemetry stream and analyzed.  They  consist of timing, energy and positional information of all the detected events. IBAS includes different triggering algorithms that search  for excesses in the light curves and in the deconvolved images, using several energy ranges and timescales. The sky coordinates of the GRBs (or other transient events) are immediately distributed.  IBAS also looks in real time for GRBs seen by the SPI/ACS, thus providing light curves with 50 ms resolution that are routinely correlated with those obtained by other satellites of the IPN  to provide GRB localizations by triangulation. IBAS is described in more detail in other articles of this volume.

Up to now (July 2019)  127 GRBs have been detected in the IBIS field of view\footnote{An updated list is available at the IBAS web pages {\it http://ibas.iasf-milano.inaf.it/IBAS\_Results.html}}.
In the following sections we first describe their global properties and then  concentrate on a few interesting cases.

\begin{table*}[ht!]
\centering
\caption{INTEGRAL GRB with redshift measurement.} 
\label{tab:redshift}
\begin{tabular}{llc}
\hline
GRB & redshift   & References   \\ 
\hline
031203  & 0.105  &   SN 2003lw \\ 
050223  & 0.5915  & \\ 
050502A & 3.793     & \\ 
050525A & 0.606    & \\ 
080603A & 1.688  &   \\ 
100518A & 4  & photometric  \\ 
120711A & 1.405  &   \\ 
140206A & 2.7  &   \\ 
141004A & 0.573  &   \\ 
160629A & 3.33  &   \\ 
161023A & 2.708  &   \\ 
181201A & 0.45 & \\
\hline
\end{tabular}
\end{table*}

\begin{figure}[ht!]
\centering
\includegraphics[scale=.30]{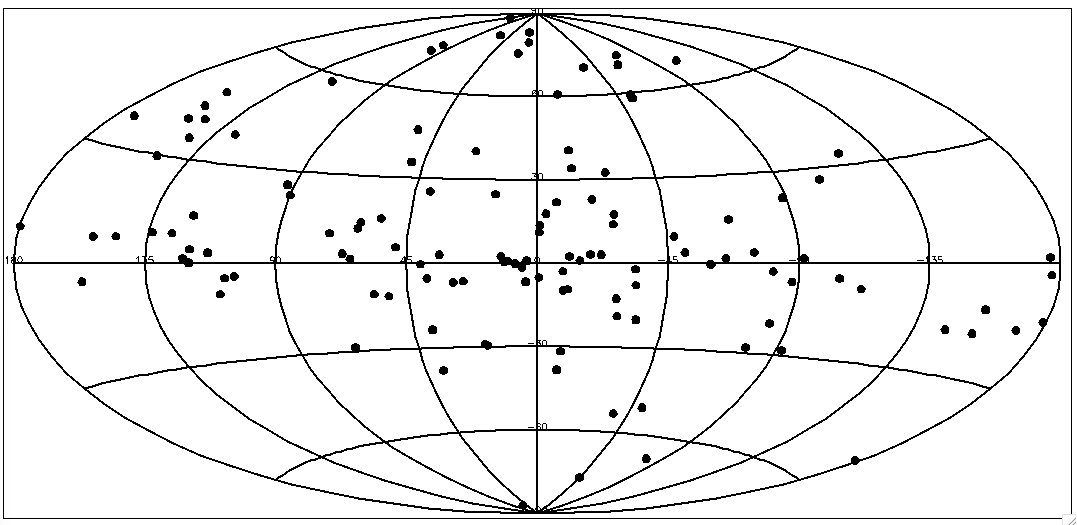}
\caption{Distribution of GRBs in Galactic coordinates.   }
\label{fig:lb}
\end{figure}

\begin{figure}[ht!]
\centering
\includegraphics[scale=.70]{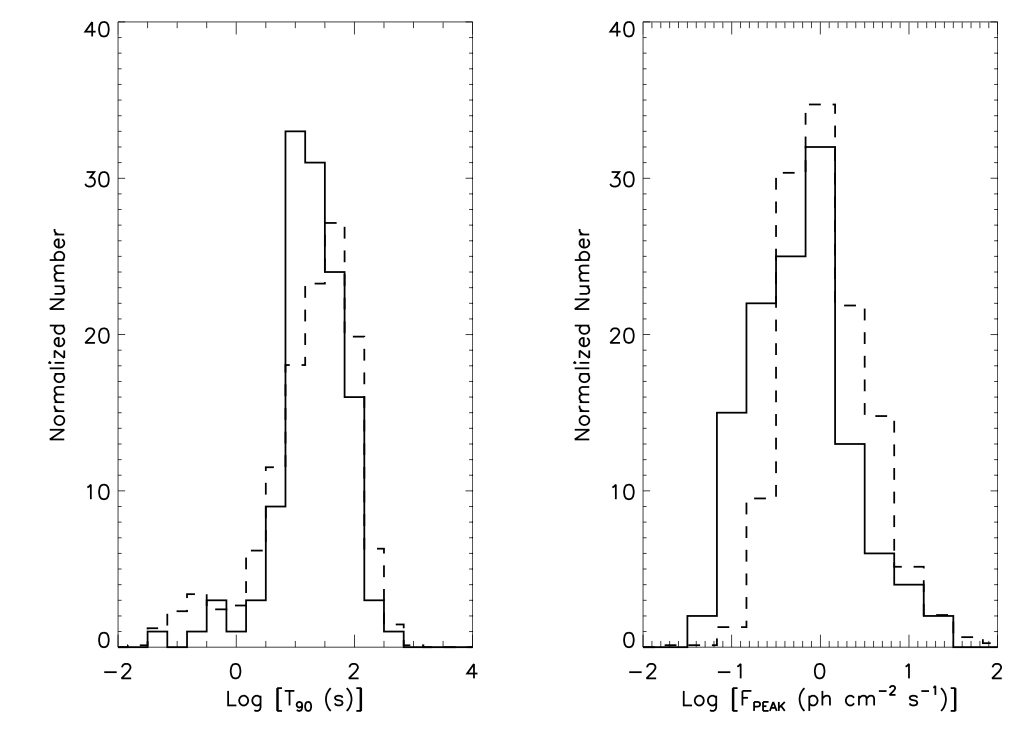}
\caption{Left: Distribution of durations of the \int\  GRBs. The dashed lines shows the distribution of Swift/BAT bursts, rescaled to the same total number of \int\  bursts. Right: Distribution of photon peak fluxes of the \int\  GRBs. The dashed lines shows the distribution of Swift/BAT bursts, rescaled to the same total number of \int\  bursts. Only long GRBs are considered; the peak fluxes are for  $\Delta T$ = 1 s and refer to the  20-200 keV energy range. }
\label{fig:t90}
\end{figure}


\subsection{Global properties}

The sky distribution of the \int\  GRBs  is shown in Fig.~\ref{fig:lb} in Galactic coordinates. The apparent anisotropy, with a large concentration of bursts at low latitudes results from the non-uniform sky coverage obtained with \int,  that has devoted many observations to Galactic sources. The unfavourable sky position of many bursts, affected by large interstellar absorption, also explains why the fraction of \int\ bursts with a measured redshift is relatively small ($\sim$10\%, see Table~\ref{tab:redshift}).

The high sensitivity of the IBIS instrument allows to sample the population of  faintest GRBs, down to peak flux values below $\sim$0.1 photons cm$^{-2}$ s$^{-1}$.  This is shown in Fig.~\ref{fig:t90}, where the distribution of peak fluxes $F_p$ is plotted together with  that of the Swift bursts\footnote{The Swift data used here and in all the other figures have been obtained from  {\it https://swift.gsfc.nasa.gov/results/batgrbcat/} }.  Excluding short GRBs,   53\% bursts of the \int\  sample have  $F_p < 1$ ph cm$^{-2}$ s$^{-1}$, compared to about 35\% for Swift.

\int\  has detected 6 short GRBs (070707,  081226B, 100703A, 110112B,  131224A,  150831A). The distribution of durations,  shown in Fig.~\ref{fig:t90}, is consistent with that of Swift GRBs.

Systematic analysis of the GRBs localized with IBAS during the first years of \int\  operations have been reported  by \cite{fol08}, \cite{VGM09}, and \cite{bos14}.  These works showed that, from the spectral point of view, the \int\ GRBs are not significantly different from those detected by other satellites.
A sample of  15 ``subthreshold'' GRBs detected with IBAS has been investigated by \cite{hig17} using follow-up observations with Swift. 

\cite{fol08} computed the spectral lags for 31 \int\ bursts. In 28 of them significant positive lags were found between the 25--50 keV and 50--300 keV light curves, indicating, as in previous works, that the low energy emission tends to be delayed compared to that of higher energy. Interestingly, it was also found that the \int\ sub-sample of faint burst ($F_p$ below $\sim$2 ph cm$^{-2}$ s$^{-1}$)   with a long spectral lag ($>$0.75 s)  shows a correlation with the supergalactic plane. The non-uniform sky distribution of these  11 GRBs was also confirmed by  \cite{VGM09},  who, besides the \int\ exposure map, took into account the reduction in sensitivity caused by the higher background in the Galactic plane  (that is particularly relevant in the case of the faintest bursts, as considered here).

The concentration of dim long-lag GRBs in the supergalactic plane found with \int\ indicates a correlation with the local structures of the Universe and supports the existence of a population of low luminosity GRBs, which is particularly interesting in the light of the recent detections of gravitational waves.

\subsection{GRB\,031203}

This relatively bright burst \cite{mer03}, one of the few \int\ GRBs with a redshift determination (z=0.106, \cite{pro04}), was spectroscopically associated with the type Ic supernova SN2003lw. Its low distance implies that GRB\,031203 has a small isotropic-equivalent energy $E_{iso}\sim10^{50}$ erg s$^{-1}$, despite the relatively large value of peak energy derived from its spectrum $E_{peak}>$190 keV \cite{saz04}. The total energy released in burst time (assuming isotropic radiation) turned out to be about three orders of magnitude lower than for typical GRBs, that makes it an outlier in the $E_{iso}$--$E_{peak}$ relation, similar to GRB\,090425. Note, that both bursts were detected only due to their relative proximity \cite{tin98,pro04}
therefore it was suggested that such low-energy bursts can be a more frequent cosmic phenomenon than ''standard'' high-energy bursts \cite{saz04,pro04}. Moreover, a detection of the expanding X-ray halo around the position of the burst with the {\it XMM-Newton} observatory on the first day after the event \cite{vau04} gave an additional interest to this burst. The origin of the soft X-ray radiation and its parameters were discussed in a number of papers \cite{vau04,tie06,wat06,saz06,ghi06}. 

The combined analysis of \int\ and {\it XMM-Newton} data is in good agreement with the fact that the bulk of the soft X-ray fluence was released during a long event ($\Delta t \simeq 100–1000$\,s), which occurred somewhat later than the GRB. Such a long emission episode could be associated both with the initial stage of the afterglow or with an additional X-ray pulse following the GRB.

Systematic studies of the soft X-ray radiation from GRBs at earlier stages, carryied out with the {\it Swift} observatory, show that powerful prolonged X-ray bursts are really observed in some cases several minutes after the GRB.  These bursts can be associated with the continuing activity of the “central engine”. It is possible that in the case of GRB\,031203, there was a similar early X-ray afterglow stage, when $\sim20$\% (if the estimate of \cite{tie06} is used) or slightly
more of the total GRB was released. This hypothesis is supported by the fact that the X-ray light curve at the late stages ($t>6$\,h) demonstrates a plateau with $\alpha \sim0.5$ \cite{wat04} and the subsequent decay with $\alpha \sim1.0$ \cite{sod04}. This scenario is also supported by the fact that the late X-ray afterglow of GRB 031203 was characterized by approximately the same spectrum ($\Gamma_X \simeq1.9$, \cite{wat04}) as the early X-ray emission from which the echo was observed.

\subsection{Polarization Results}
\label{sec:grb_pola}
Despite the recent \textbf{progress} in the GRB field obtained mainly thanks to the \textit{Swift} and \textit{Fermi} satellites \citep[see e.g.][]{gehrels09,zhang11}, the nature of their prompt emission is still not clear. In particular, the precise content of the jet, and especially its magnetization, as well as the details of the mechanism leading to the $\gamma$-ray emission are still not completely elucidated. Models range from unmagnetized fireballs where the observed emission could be produced by electrons accelerated in internal shocks propagating within the relativistic outflow ($\Gamma > 100$) \cite{rees94}, to pure electromagnetic outflows where the radiated energy comes from magnetic dissipation \cite{lyutikov06}. Intermediate cases with mildly magnetized outflows are of course also possible \cite[e.g.][]{spruit01}.
Even in the case of an unmagnetized fireball, a local magnetic field in the emission region, generated by the shocks, is necessary if the dominant process is synchrotron radiation from relativistic electrons. 


In this article we will focus on the observational results obtained by \int\ and we \textbf{point} the reader interested in the details of the theoretical modeling of prompt and afterglow polarimetric measurements to the review by \cite{covinogotz16}.

The measurement of polarization during the prompt phase of GRBs has always been challenging. This is mainly due to the fact that no wide field $\gamma$-ray polarimeter with a large effective area has yet been flown, and that many of the measurements attempted to date have been performed with instruments which have some polarimetric capabilities, but have not an explicit polarimetric oriented design. In addition, the GRB prompt emission is very limited in time, mostly less than $\sim$100 s, and hence in spite of the high average flux of GRBs, the total number of collected photons is often too limited to derive statistically stringent limits for polarization.

It is worth \textbf{remembering} that before the \int\ results, some early attempts to measure GRB polarization have been performed. For instance \cite{coburn03} used RHESSI observations of GRB 021206 to try to measure its polarization. In this case, as for all the $\gamma$-ray polarimeters, the polarization dependency of the differential cross section for Compton scattering was exploited to constrain the degree and angle of polarization of the incident radiation:

\begin{equation}
\frac{d\sigma}{d\Omega} = \frac{r_{0}^{2}}{2}\left(\frac{E^{\prime}}{E_{0}}\right)^{2}\left(\frac{E^{\prime}}{E_{0}}+\frac{E_{0}}{E^{\prime}}-2 \sin^{2}\theta \cos^{2}\phi \right)
\label{eq:cross}
\end{equation}

where $r_{0}^{2}$ is the classical electron radius, $E_{0}$ the energy of the incident photon, $E^{\prime}$ the energy of the scattered photon, $\theta$ the scattering angle, and $\phi$ the azimuthal angle relative to the polarization direction. 
Linearly polarized photons scatter preferentially perpendicularly to the incident
polarization vector. Hence by examining the angles of scattering of the photons among different detectors of a same instrument, one can in principle derive the degree and angle of linear polarization of the incident photons.

Using this technique \cite{coburn03} reported a high level of linear polarization ($\Pi$=80$\pm$20\%) at a high level of confidence ($>$ 5.7$\sigma$) for GRB 021206. 
However subsequent re-analyses of the same data performed by different groups \citep{rutledge04,wigger04} could not confirm this result.

Other early attempts to measure the polarization of GRBs include the ones performed by \cite{willis05}, who used the Earth atmosphere as a scatterer and BATSE as a detector. They reported the evidence of high linear polarization in two GRBs (GRB 930131 and 960924), $\Pi >$35\% and $\Pi >$50\% respectively. However due to the complexity of their method, they could not statistically constrain their result.

\subsubsection{\int\ Results}

At the time of its discovery GRB 041219A \citep{mcbreen06,gotz11} was among the top 1\% in terms of GRB fluence. This prompted different groups to try to measure its polarization with \int. The first attempt was performed using SPI, and exploiting the scatterings among its individual hexagonal Ge detectors. \cite{kalemci07} reported a high level of polarization for this GRB ($\Pi=98\pm33\%$), but could not constrain the systematics of their measurements. Using the same data set but a more sophisticated analysis technique, where the multiple event scatterings in the  SPI spectrometer have been compared to a GEANT4 Monte Carlo simulation predicted response to a polarized source flux, \cite{mcglynn07} were able to measure the degree of linear polarization over the brightest pulse of the GRB (lasting 66 s) to $\Pi=63^{+31}_{-30}\%$ and the polarization angle to $P.A.=70^{\circ}$ $^{+14}_{-11} $ . 

GRB 041219A was also observed by IBIS. Thanks to its two superposed pixellated detection planes -- ISGRI \citep{isgri} and PICsIT \citep{picsit}, IBIS can be used as a Compton Polarimeter. For the details on how the polarization can be measured by SPI and IBIS, see Section \ref{sec:pola}.

Generally, following the principle expressed in Eq. \ref{eq:cross}, and examining the scatter angle distribution of the detected photons in the two detectors, expressed as

\begin{equation}
N(\phi)=S[1+a_{0}\cos 2(\phi-\phi_{0})],
\label{eq:azimuth}
\end{equation}

one can derive the polarization angle, $PA =  \phi_{0} - \pi /2 + n \pi$, and the polarization fraction $\Pi= a_{0}/a_{100}$, where $a_{100}$ is the amplitude expected for a 100\% polarized source.

\begin{figure*}[ht!]
\centering
\includegraphics[width=13cm]{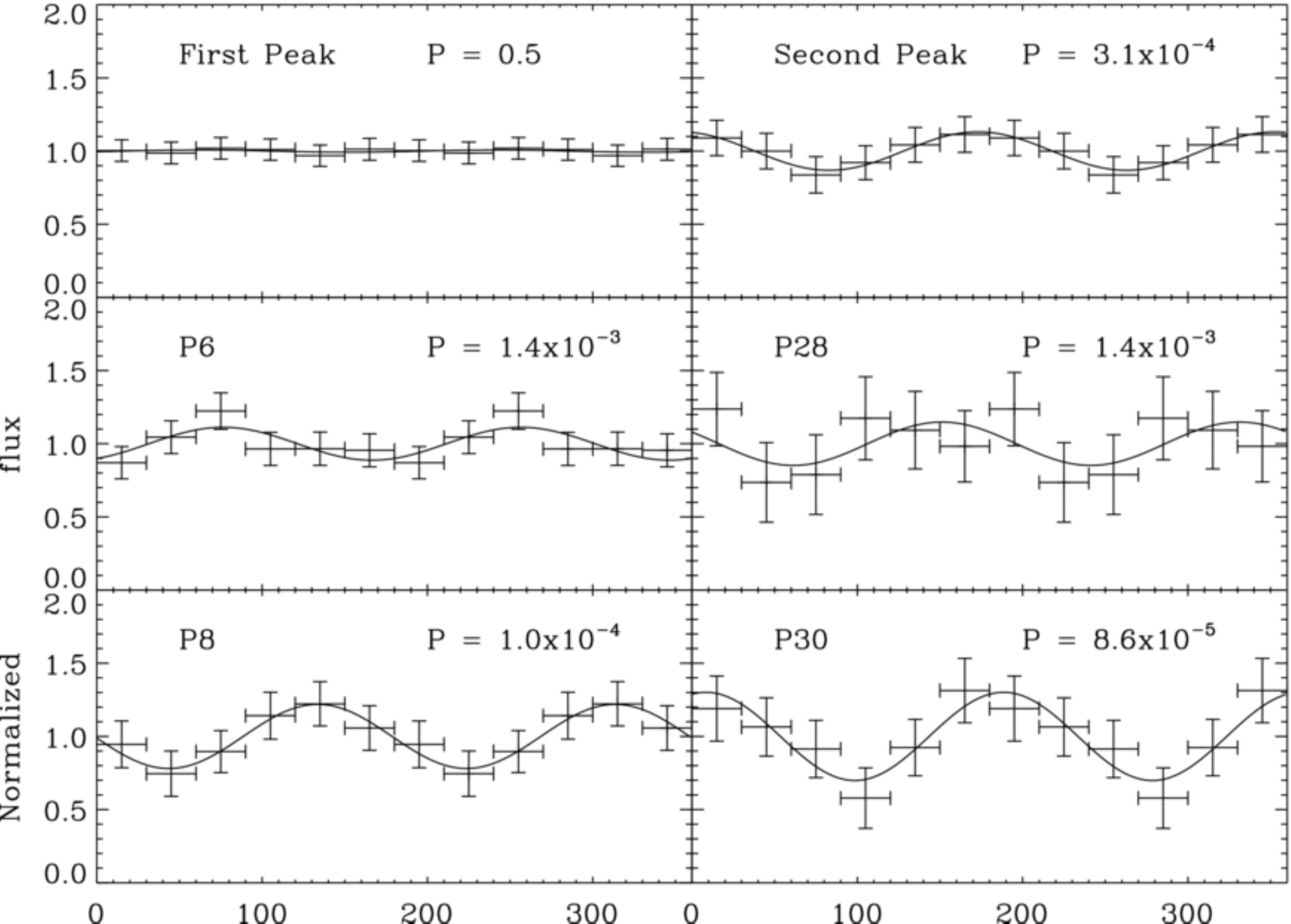}
\caption{Polarigrams of the different time intervals that have been analysed for GRB 041219A (see Table \ref{tab:pola}). For comparison purposes, the curves have been normalized to their average flux level. The crosses represent the data points (replicated once for clarity) and the continuous line the best fit. For each polarigram the probability, $P$, is shown that the polarigram measured corresponds to an un-polarized ($\Pi<$1\%) source. From \cite{gotz09}.}
      \label{fig:lc}
\end{figure*}

\begin{figure*}[ht!]
\centering
\includegraphics[width=6.5cm]{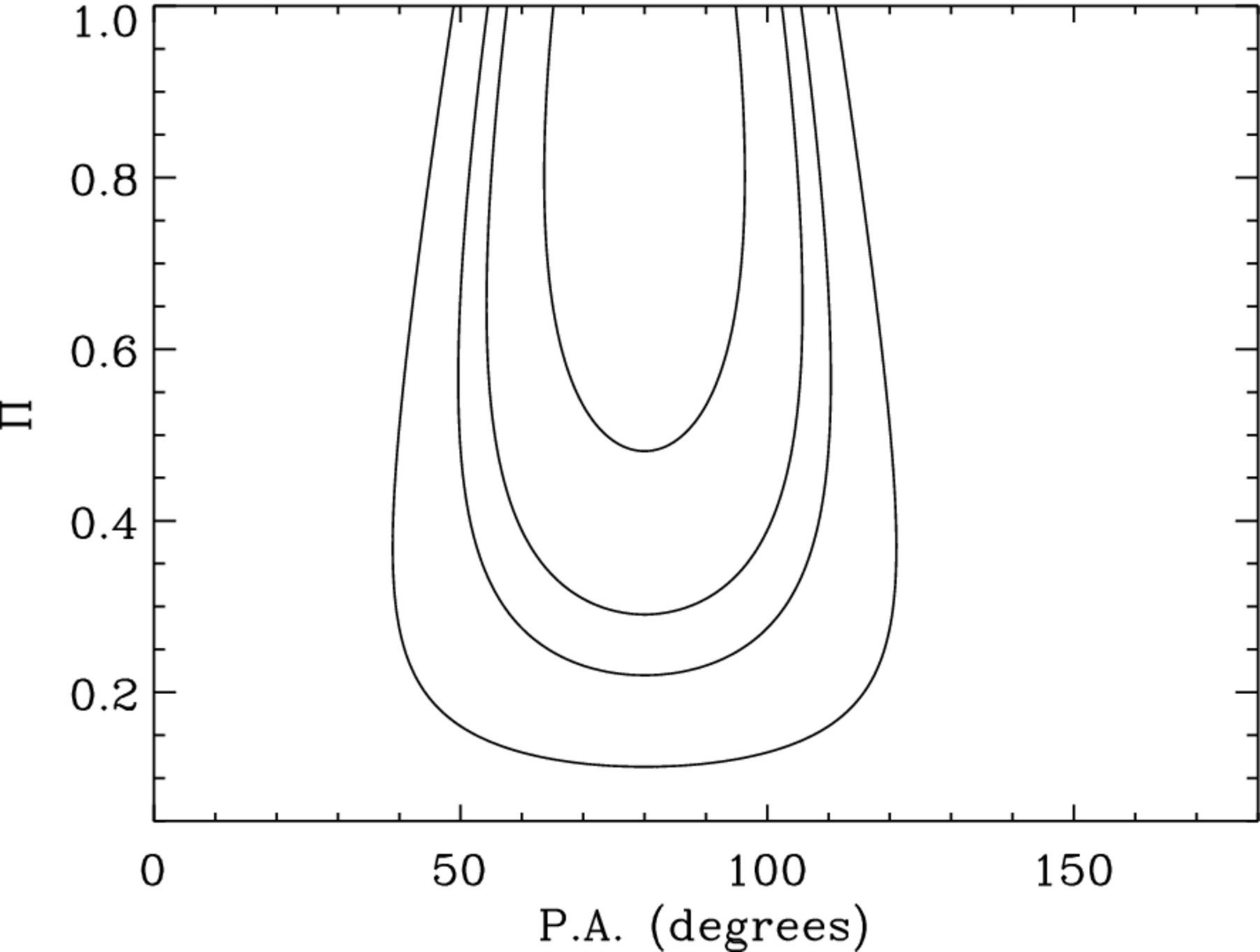}
\includegraphics[width=6.5cm]{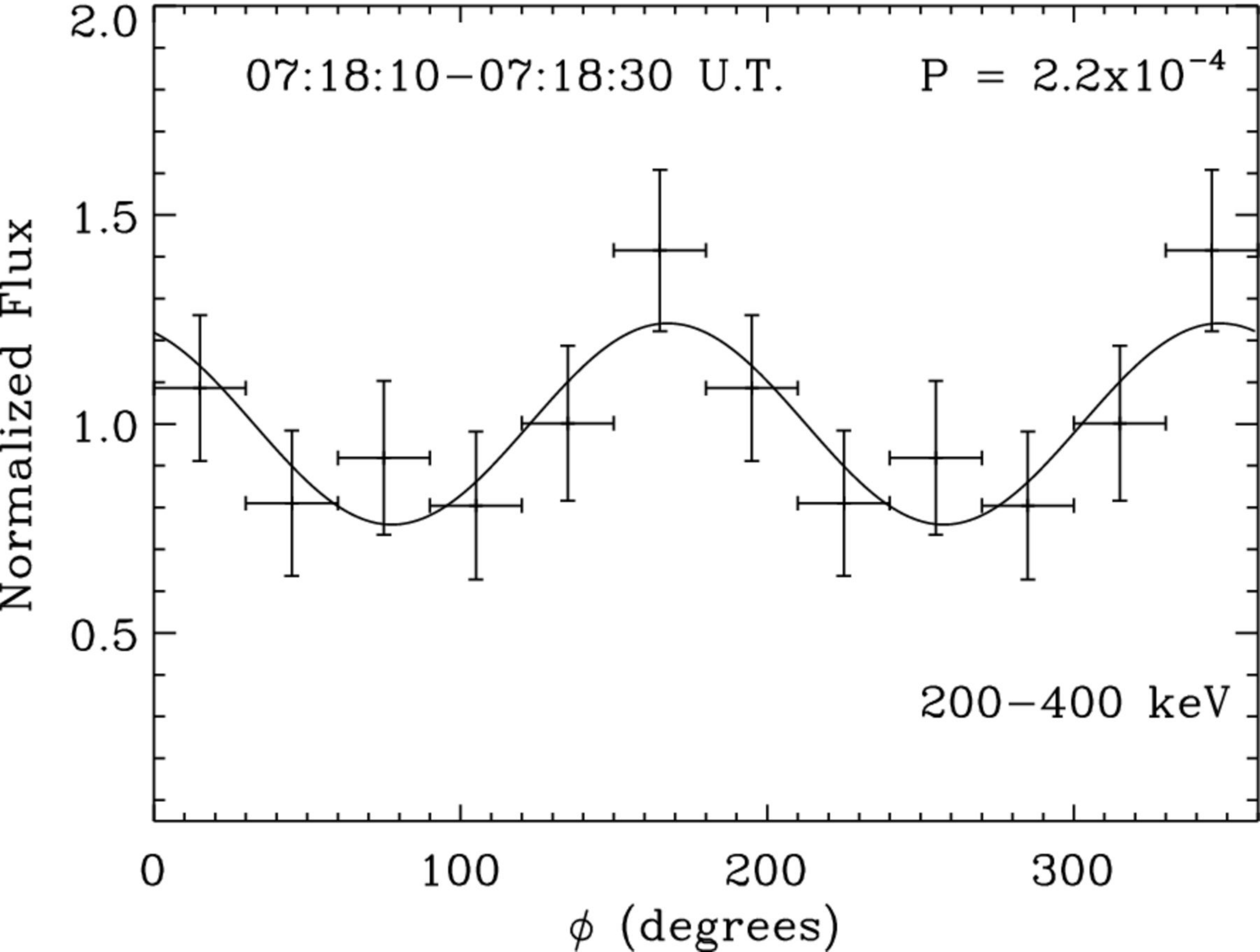}
\caption{Left: The 68, 90, 95 and 99\% confidence contours for the $\Pi$ and PA parameters. Right:  Polarigram of GRB 140206A in the 200--400 keV energy band. The crosses represent the data points (replicated once for clarity) and the continuous line the fit done on the first six points using equation \ref{eq:azimuth}. The chance probability P of a non-polarized ($<$1 \%) signal is also reported. From \cite{gotz13}.}
      \label{fig:140206A}
\end{figure*}

\begin{table*}[ht]
\caption{Polarization results for the different time intervals for GRB 041219A. From \cite{gotz09}}
\begin{center}
\begin{tabular}{cccccc}
\hline
Name & T$_{start}$ & T$_{stop}$ & $\Pi$ & $PA$ & Image\\
& U.T.  & U.T. & \% & degrees ($^{\circ}$) & SNR \\
\hline
First Peak & 01:46:22 & 01:47:40 & $<$4 & -- & 32.0\\
Second Peak & 01:48:12 & 01:48:52 & 43$\pm$25 & 38$\pm$16 & 20.0\\
P6 & 01:46:47 & 01:46:57 & 22$\pm$ 13& 121$\pm$17 & 21.5\\
P8 & 01:46:57 & 01:47:07 &65$\pm$26 & 88$\pm$12 & 15.9\\ 
P9 & 01:47:02 & 01:47:12 &61$\pm$25 & 105$\pm$18 & 18.2\\
P28 & 01:48:37 & 01:48:47 &42$\pm$42 & 106$\pm$37 & 9.9\\
P30 & 01:48:47 & 01:48:57&90$\pm$36 & 54$\pm$11 & 11.8\\
\hline
\end{tabular}
\end{center}
Errors are given at 1 $\sigma$ c.l. for one parameter of interest.
\label{tab:pola}
\end{table*}

Using the same method polarization could be measured for two other GRBs (061122 and 140206A), with IBIS \citep{gotz13,gotz14}, see Tab. \ref{tab:polasummary}, but no time-resolved analysis could be performed due to the limited statistics, making of GRB 041219A the only GRB for which a time variable polarization signal could be measured to date with IBIS. We note that for GRB 041219A the polarization values obtained by SPI over the 12 brightest s of the GRB ($\Pi=68\pm29\%$, $P.A.=70^{\circ}$ $^{+14}_{-10}$) are compatible with the ones obtained by IBIS on the equivalent P8 (see Table \ref{tab:pola}) interval ($\Pi=65\pm26\%$, $P.A.=88^\circ \pm 12^\circ$). Concerning 061122 a SPI polarization measurement by \cite{mcglynn09} has also been published, and the IBIS and SPI results
are compatible if one considers the 90\% c.l. IBIS lower limit ($\Pi >$ 33 \%) and the
SPI 95\% c.l. upper limit ($\Pi <$ 60 \%).

\begin{table*}[ht!]
\centering
\caption{Summary of recent GRB polarization measurement by IBIS/SPI.} 
\label{tab:polasummary}
\begin{tabular}{cccccc}
\hline
GRB & $\Pi$   & Peak energy & Fluence & Energy Range & Instrument\\ 
& (68\% c.l.) & (keV) & (erg cm$^{-2}$) &  &\\
\hline
041219A & 65$\pm$26\% & 201$^{+80}_{-41}$ & 2.5$\times 10^{-4}$ & 20--200 keV & IBIS, SPI\\ 
061122 & $>$60\% & 188$\pm$17 & 2.0$\times 10^{-5}$ & 20--200 keV & IBIS, SPI\\ 
140206A & $>$48\% & 98$\pm$17 & 2.0$\times 10^{-5}$ & 15--350 keV & IBIS\\ 
\hline
\end{tabular}
\end{table*}

Although all these measures, taken individually, have not a very high significance ($\sim$2.5 $\sigma$), they indicate that GRBs may indeed be emitters of polarized radiation, since compatible results have been obtained with two different instruments (SPI and IBIS) on at least two GRBs.

As discussed e.g. in \cite{covinogotz16}, the measure of a high level of linear polarization, as well as its variability, point towards an interpretation where synchrotron radiation is emitted from shock accelerated electrons in a relativistic jet with a magnetic field transverse to the jet expansion. The coherence of the magnetic field geometry does not need to hold over the entire jet, but only over a small portion of it, since, due to relativistic effects, the observer can see only a region of the jet whose angular size is comparable to 1/$\Gamma$, $\Gamma$ being the Lorentz factor of the relativistic outflow. If the radiating electrons are accelerated in internal shocks \citep{rees94,kobayashi97,daigne98}, then the Lorentz factor is necessarily varying in the outflow, which can explain the variability of the polarization from one pulse to the other \citep{granot03a,granot03b,nakar03}, as observed in GRB 041291A and GRB 100826A (see later). This is not the only possible interpretation, and other models that predict locally coherent magnetic fields, like fragmented fireballs (shotguns, cannonballs, sub-jets) can produce highly polarized emission, with a variable polarization angle \citep[e.g.][]{lazzati09}. Even in the case of \textbf{magnetically dominated} jet models a change in polarization angle is sometimes expected \citep{chang14}. In addition, different emission mechanisms can not be completely excluded at this time, implying random magnetic fields and peculiar observing conditions, like e.g. inverse Compton scattering \citep{lazzati04}. \textbf{Better quality data, for a larger sample of bursts, will be required to exclude certain models on the basis of statistical arguments.}

\subsubsection{Comparison to GAP and POLAR results}

We briefly report here the results from two other instruments, GAP and POLAR. Unlike SPI and IBIS, they have been designed and and optimized for being polarimeters, and both have confirmed the polarized nature of Gamma-Ray Burts.

The GAP experiment is a Japanese Compton polarimeter launched in May 21st 2010, on-board the IKAROS solar sail \citep{yonetoku}. It is composed of an assembly of two coaxial detectors, made respectively of a plastic scintillator and CsI crystals. Its polarimetric capabilities have been tested on ground in polarized beams to calibrate systematic effects. In 2010,  GRB 100826A has been observed by GAP in the 70 to 300 keV energy band. Its modulation curve was fitted with a Monte-Carlo model of the experiment \citep{yonetoku}, and a (time variable) polarized signal was marginally observed with a significance of $2.9 ~\sigma$. Furthermore, in 2011, the GAP instrument measured the polarization of two more Gamma-Ray Bursts, GRB 110301A and GRB 110721 A. 



The POLAR instrument is a Swiss/Chinese mission dedicated to the Gamma-Ray Burst polarization measurements. It has been launched on September 15th, 2016 and placed on the Chinese space station Tiangong 2 \citep{produit}. This Compton telescope is made by 1600 bars of plastic scintillators read out by photomultiplier in an array of $40 \times 40$. Its conception ensures a very wide field of view, one-third of the sky, and very good polarimetric capabilities ($a_{100} \approx 35 \%$). Due to a failure of its power distribution unit, the telescope was \textbf{switched} off in 2017.


The POLAR telescope has measured 55 Gamma-Ray Bursts light curves. Among these, five were bright enough in the 10 -- 1000 keV energy range to enable a polarization analysis. This work has revealed that, on average over the five GRB duration, the polarization fraction is rather low, around $10 \%$. However, when considering time resolved analysis for the brighter bursts, like GRB 170114A, polarization is measured to be as high as $28 \%$. This may indicate that the $\gamma$-ray light \textbf{is mildly} polarized, but as the polarization angle changes during the burst,  on average, we measure a low polarization degree  \citep{Zhang}.  

\subsubsection{Constraints on Lorentz Invariance Violation}
\label{sec:liv}

An interesting aspect related to measurements of linear polarization of GRBs and in general of cosmological sources is the \textbf{possibility} to constrain Lorentz Invariance Violations (LIV). A possible experimental test for such violation is to measure the helicity dependence
of the propagation velocity of photons \citep[see e.g.][and references
therein]{laurent11a}.

In general, it is possible to describe light as composed of two independently propagating constituent waves, each possessing a polarization and a velocity. Certain forms of relativity violations cause light to experience birefringence, a change in polarization as it propagates. The changes grow linearly with distance travelled, so birefringence over cosmological scales offers a sensitive probe for relativity violations. Searches for this vacuum birefringence using polarized light emitted from sources at cosmological distances yield some of the sharpest existing tests of relativity \citep{firstliv}.


 The light dispersion relation is given in this
case by

\begin{equation}
\omega^{2}=k^{2}\pm\frac{2\xi k^{3}}{M_{Pl}}\equiv\omega^{2}_{\pm}
\label{eq:dispersion1}
\end{equation}

where $E=\hbar\omega$, $p=\hbar k$, $M_{Pl}$ is the Planck Mass, and the sign of the cubic term is determined by the chirality (or circular polarization) of the photons, which leads to a rotation of the polarization during the propagation of linearly polarized photons. This effect is known as vacuum birefringence.

Equation \ref{eq:dispersion1} can be approximated as follows

\begin{equation}
\omega_{\pm}=\vert k \vert \sqrt{1\pm\frac{2\xi k}{M_{Pl}}}\approx\vert k\vert(1\pm\frac{\xi k }{M_{Pl}})
\label{eq:dispersion2}
\end{equation} 

where $\xi$ gives the order of magnitude of the effect. In practice some quantum-gravity theories \citep[e.g.][]{myers03} predict that the polarization plane of the electromagnetic waves emitted by a distant source rotates by a quantity $\Delta\theta$ while the \textbf{wave} propagates through space, and this as a function of the energy of the photons, see Eq. \ref{eq:rotation}, where $d$ is the distance of the source:

\begin{equation}
\Delta\theta(p)=\frac{\omega_{+}(k)-\omega_{-}(k)}{2}\ d\approx\xi\frac{k^{2}d}{2M_{Pl}}
\label{eq:rotation}
\end{equation}

As a consequence the signal produced by a linearly polarized source, observed in a given energy band could vanish, if the distance is large enough, since the differential rotation acting on the polarization angle as a function of energy would in the end add \textbf{oppositely} oriented polarization vectors, and hence in a net un-polarized signal.
\textbf{Since this effect is extremely small, being} inversely proportional to the Planck Mass ($M_{Pl}\sim$2.4$\times$10$^{18}$ GeV), the observed source needs to be at cosmological distances. The simple fact to detect the polarization signal from a distant source, can put a limit to such a possible violation. This experiment has been performed by \cite{laurent11a}, \cite{toma12}, and \cite{gotz13} making use of the prompt emission of GRBs. Indeed, since GRBs are at the same time at cosmological distances, and emitting at high energies, their polarization measurements are highly suited to measure and improve upon these limits.

By using the distance measured from the afterglow absorption spectrum of GRB140206A (23 Gpc) \cite{gotz14} obtained

\begin{equation}
\xi < \frac{2 M_{Pl}\Delta\theta(k)}{(k_{2}^{2}-k_{1}^{2})\ d}\approx 1\times 10^{-16},
\label{eq:xi}
\end{equation}

which is the most stringent limit to date using this technique.

\section{Polarization of Hard X-ray Sources}
\label{sec:pola}

\subsection{The INTEGRAL polarimetric capabilities}

Both \int\ main telescopes, IBIS and SPI, can be used as polarimeters. In this section we
describe briefly how the polarization measurements are performed for both instruments.


\subsubsection{SPI as a polarimeter}

The polarization capabilities of the SPI instrument rely on the detection of multiple events (when a photon deposits energy in two or more detectors). They  thus start above $\sim$ 100 keV. In short, the data analysis for this specific purpose is based on  GEANT 4 Monte-Carlo simulations of the SPI response to polarized emissions. These simulations are performed for 18 angles, from 0 to 170$^{\circ}$, by step of 10$^{\circ}$. The un-polarized simulation data complete  the predicted data set, and, by combination with the polarized data, make it possible to simulate any polarization fraction. The comparison of the predicted and observed data allows us to determine the polarization angle and polarization fraction which give the best description of the source emission. The first tools have been presented in \cite{mcglynn09} and applied to bright GRBs, then, to the Crab nebula off-pulse emission \citep{dean08}. Later, an improved version of the Monte-Carlo simulation tools has been developed and used to analyzed the total (Nebula + pulsar) Crab emission by  \cite{chauvin13} .



\subsubsection{IBIS as a polarimeter}
Thanks to its two position sensitive detectors ISGRI (made of CdTe crystals and sensitive in the 15--1000 keV energy band), and PICsIT (made of CsI bars and sensitive in the 200 keV--10 MeV energy band), the IBIS telescope has been also used as a Compton polarimeter to study many compact objects. The procedure to measure the polarization is described in \cite{forot08}. \textbf{It allowed systematic effects to be controlled} and to successfully detect a polarized signal from the Crab nebula. To derive the source flux as a function of $\phi$, the Compton photons were divided into 6 bins of 30$^{\circ}$ as a function of the azimuthal scattering angle. The chance  coincidences (i.e. photons interacting in both detectors, but not related to a Compton event), have been subtracted from each detector image also following the procedure described in \cite{forot08}. The derived detector images were then deconvolved to obtain sky images, where the flux of the source in each azimuthal bin was measured by fitting the instrumental  PSF to the source peak.  The azimuthal profiles, called polarigrams, (see Fig. \ref{fig:integral_Crab}) were fitted using a least squares technique to derive $a_{0}$ and $\phi_{0}$. \newline

\subsection{INTEGRAL observations of the Crab pulsar and nebula polarization}

\subsubsection{IBIS result on the Crab nebula}

\begin{figure}[ht!]
\centering
\includegraphics[scale=0.45]{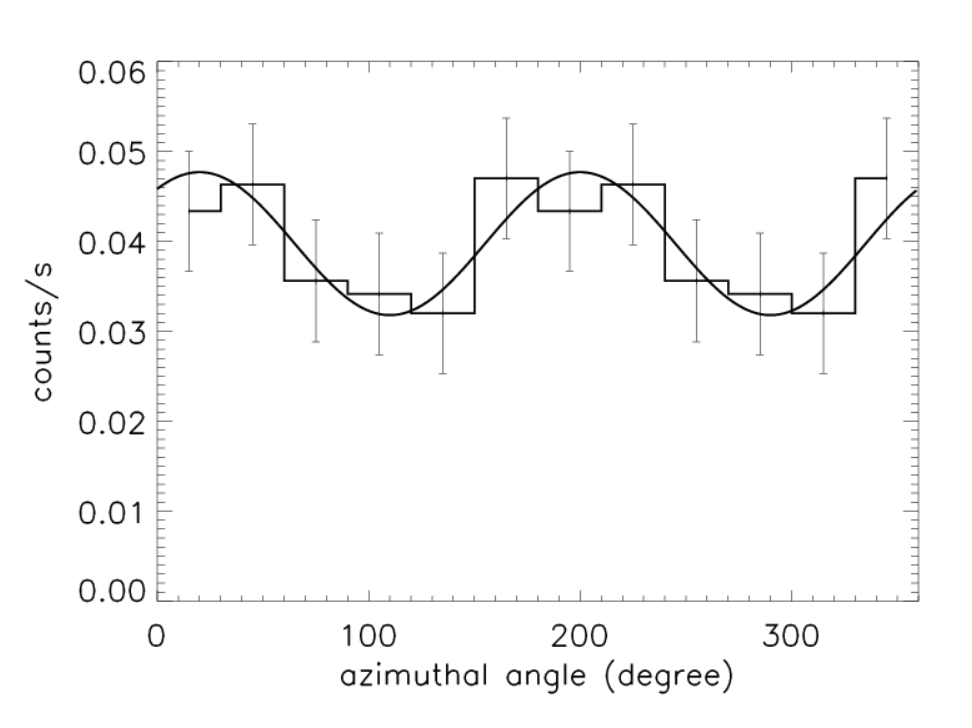}
\caption{ Phase-averaged polarigram of the Crab pulsar/nebula obtained by \int\ IBIS in fall 2015 in the 300--800 keV energy band.}
\label{fig:integral_Crab}
\end{figure}

The Crab nebula is regularly observed since the \int\ launch for calibration purposes. Polarization results obtained with IBIS were first published in \cite{forot08} and \cite{moran16} covering the period 2002--2014. 
In Figure \ref{fig:integral_Crab}, we show the polarigram obtained with IBIS using data obtained between 2015, August 18th and October 17th for a total exposure time of 500 ksec. Optical polarimetric Crab observations were also performed at the same epoch at 
the William Herschel Telescope (WHT), using the GASP polarimeter \citep{gouiffes16}. 
Several observations have demonstrated that the Crab polarization angle has shifted during the 2003--2015 period, jointly in optical and $\gamma$-rays (see Table \ref{table_int}). 

\begin{table}[ht!]
\centering
\caption {\int\ IBIS  polarization parameters in the 300--450 keV band compared to optical measurements}
\vspace{0.3cm}
\begin{tabular}{ccccc}
\hline
Date & PA ($^o$) & PF & Optical PA ($^o$) & Reference\\
\hline
2003--2007 & 115  $\pm$ 11 & 96 $\pm$ 34 & 109.5  $\pm$ 0.7 & \citep{forot08} \\
2012--2014 & 80  $\pm$ 12 & 98 $\pm$ 37 & 85.3  $\pm$ 1.4 & \citep{moran16} \\ 
2015 & 125  $\pm$ 15 & 89 $\pm$ 28 & 130 $\pm$ 1.0 &\citep{gouiffes16} \\
\hline
\end{tabular}
\label{table_int}
\end{table}

\subsubsection{SPI result on the Crab nebula}
\begin{figure}
\includegraphics[scale=0.5]{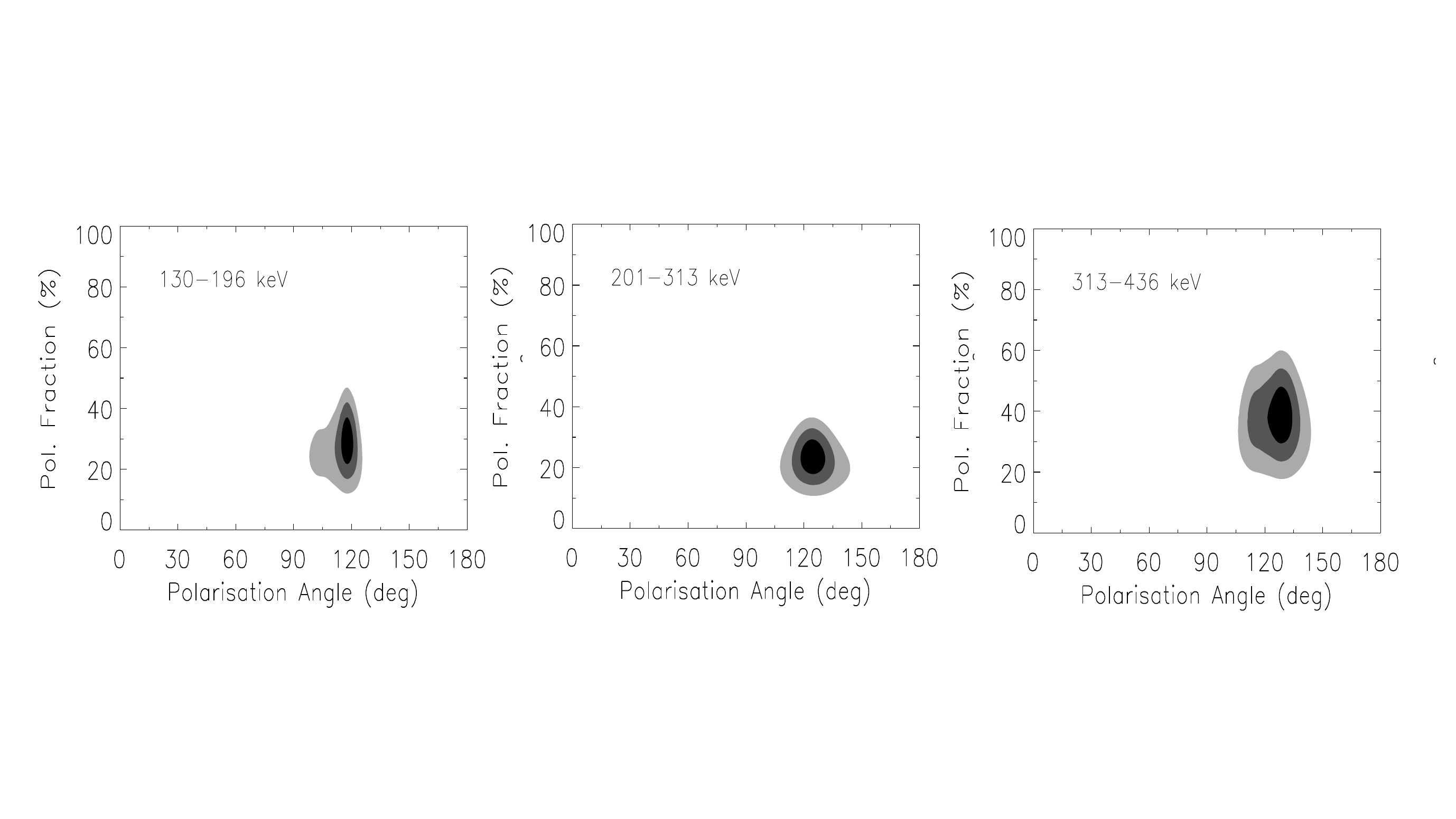} 
\caption{Crab Polarization parameters in 3 energy bands (SPI data,  2003-2018 observations).}\label{CrabSPIvsE}
\end{figure}

 \cite{dean08} reported on the first detection of the polarization of the Crab pulsar in the Hard X-ray domain. They found that the off-pulse emission
is highly polarized (46 $\pm  10 \%$), with an electric vector aligned with the spin axis of the pulsar (124$^{\circ}$). Considering the total (pulsar + nebula) Crab emission, \cite{chauvin13} determined a lower averaged polarization fraction ($28 \pm 6\%$),  with a similar polarization angle value ($117^{\circ} \pm 9^{\circ} $). Recently, \cite{crab19} analyzed 16 yr of Crab observations by \int\ SPI. The parameters determined over the total dataset, for the pulsar + nebula emission are  PA=$120^{\circ} \pm 6^{\circ} $  and PF=$24 \pm 4\%$, in agreement with the previously reported values. To study the evolution of the polarization properties with time, the dataset has been split into four periods, analyzed separately. The polarization parameters are found to be consistent within uncertainties for all periods. Similarly, these parameters do not change with energy, from 130 to 426 keV (see fig.\ref{CrabSPIvsE}).

\subsection{Cygnus X--1 polarization}

The X-ray binary Cygnus X--1 is the first Galactic source known to host a black hole with a  mass of $14.8 \pm 1.0$  M$_\odot$ \citep{orosz11}. The  donor 
star is the high-mass O9.7 Iab star HDE 226868. The system is located at a distance of d = $1.86 \pm 0.12$ kpc from Earth with an orbital period of 5.6 days.

Cygnus X--1 is mainly observed in two main spectral states, the Low Hard State (LHS), characterized by a Comptonized thermal emission with a temperature around 50 keV and the High Soft State (HSS), dominated by a bright X-ray thermal emission coming most probably from the accretion disk. From 2002 to 2011 Cygnus X--1 was predominantly in the
LHS. It then transited in the HSS at the end of 2011 and remained mainly in this state 
until the end of 2016, for one of its longest period of HSS. It sometimes undergoes
partial (“failed”) transitions from the LHS to the HSS, and can be found in a transitional or intermediate state. The detection of compact relativistic jets in the LHS \citep{stirling01} places Cyg X--1 in the family of microquasars.

The presence of a MeV tail in the Low Hard State, first detected by COMPTEL \citep[e.g.][]{comptel}, has recently been confirmed by the two main instruments onboard \int\ SPI \citep{jourdain12}, and IBIS \citep{rodriguez15}.
Analysing SPI and IBIS data,  \cite{jourdain12}, and \cite{laurent11b}, respectively,
have been first to show that the  Cygnus X--1 emission above 400 keV is polarized at a level of about $70\%$. They obtained a $20\%$ upper limit on the degree of polarization at lower energies. Both teams suggested that the polarized emission was due to synchrotron emission coming from a compact jet. 

\cite{rodriguez15} and \cite{Cangemi19} performed a state-resolved analysis of Cygnus X--1 polarization and showed that the high-energy tail above 400 keV is detected by \int\ either in LHS or HSS. They measured polarisation in the LHS and found results in agreement with previous works, however indicating that the polarization angle is possibly time-varying on month/years timescale. No polarization was detected up to now in the HSS with a upper limit around $50 \%$.

The measure of polarized emission and of its energy dependency gives strong constraints on our understanding of accretion and ejection processes. It helps to distinguish between the different proposed emitting media (Comptonization corona versus synchrotron jets) in microquasars and provide important clues to the composition, energetics and magnetic field of the jet.

\subsection{Observation of V404 Cygni polarization}

The black hole candidate and microquasar V404 Cygni went into outburst in 2015. During this episode it was the brightest X-ray source in the entire celestial sky reaching 
peak fluxes of about 50 times the Crab Nebula in the standard IBIS range \citep[typically 20--40 keV,][]{rodriguez15b}.  

\begin{figure}[ht!]
\centering
\includegraphics[scale=1.0]{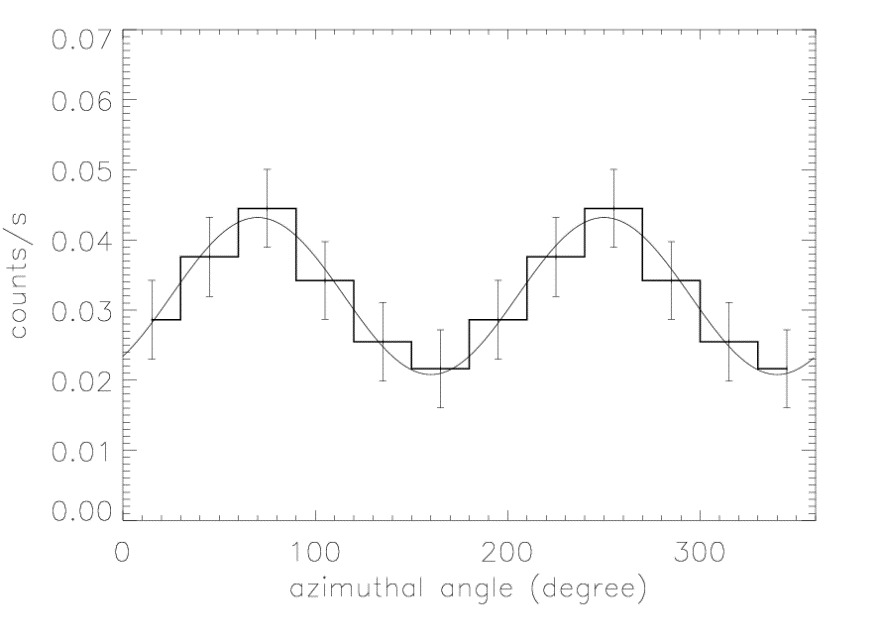}
\caption{ Phase-averaged polarigram of V404 Cygni obtained by \int\ IBIS in fall 2015 in the 300--450 keV energy band. The polarization angle is around $160^\circ$.}
\label{fig:integral_v404}
\end{figure}

The exceptional bright flaring
activity then source showed  offered a unique opportunity to study into details the polarimetric characteristics of the source. \cite{Laurent16}  have studied the V404 Cygni spectral and polarimetric properties using a new data analysis software dedicated to very bright sources, where telemetry losses occur. They have computed the V404 Cygni polarigrams, each time taking into account all data recorded during one \int\ revolution. The study of the 10 revolutions dedicated to the source in June 2015 (see Table \ref{tab:v404}) shows quite surprisingly that polarization is detected only during rev. 1555. Indeed, the source was even brighter in rev. 1557 but we detected no polarization signal at that time (PF $< 17\%$). This may imply that the signal is truly non-polarized or that the polarization angle varied during these revolutions. The polarigram obtained during rev. 1555 is shown on Figure \ref{fig:integral_v404}; the polarization parameters are PA = $160^\circ ~\pm~ 15 ^\circ$ and PF = $95 ~ \pm~ 35 \%$. 

\begin{table*}[ht!]
\caption{: \int\ IBIS V404 Cygni polarization study in the [400-2000] keV band (June 2015). If we sum all events over a revolution, a clear signal is only visible for rev. 1555. No signal is detected when the source is the brightest (rev. 1557).} 
\label{tab:v404}
\vspace{0.3cm}
\hspace{-0.8cm}
\begin{tabular}{ccccccccccc}
\hline
Rev. & 1554&1555&1556&1557&1558&1559&1560&1561&1562&1563\\ 
number \\
\hline
Exp. time&170&	149&	84&	153&	163&	171&	93&	165&	174&	117\\
(ks)\\
\hline
Compton& 8.8&	14&	4.6&	28&	-0.4&	-0.2&	-0.05&	1.26&	1.75&	1.09\\
S/N \\
\hline
Polarization& No&	Yes&	No&	No&	No&	No&	No&	No&	No&	No\\
detected ? \\
\hline
\end{tabular}
\end{table*}

These results are consistent with the near infrared (NIR) polarization detection with a polarization angle of $171 ^\circ$ \citep{shahbaz16}. It is worth noting however that the IBIS measurements occured, according to this preliminary analysis, three days before the NIR measurements, and that no polarization signal contemporaneously to the NIR data (around \int\ revolution 1556) was detected \citep{Laurent16}.

\section{Acknowledgements}
DG, CG, PL \& JR acknowledge partial funding from the French Space Agency (CNES). AL acknowledges support from Russian Foundation for Basic Research, grant 17-52-80139. DG acknowledges the financial support of the UnivEarthS Labex program at Sorbonne Paris Cit\'e (ANR-10-LABX-0023 and ANR-11-IDEX-0005-02).

\end{document}